\documentclass[twocolumn,floatfix,aps,superscriptaddress]{revtex4}
\usepackage{subfigure,dcolumn,jabbrv}
\usepackage{graphicx}% Include figure files
\usepackage{dcolumn}% Align table columns on decimal point
\usepackage{bm}% bold math
\usepackage{CJK}
\usepackage{url}
\usepackage{color}
\usepackage{booktabs}
\usepackage{natbib}

\usepackage{amsmath,bm}
\usepackage[utf8]{inputenc}
\usepackage[british,UKenglish,USenglish,american]{babel}
\usepackage{bbm}
\usepackage{amssymb}
\usepackage{amsmath}
\usepackage{hyperref}
\usepackage{cleveref}
\usepackage{tikz}\usetikzlibrary{shapes.geometric, arrows}
\usepackage{makecell}
\tikzstyle{startstop} = [rectangle, rounded corners, minimum width=3cm, minimum height=1cm,text centered, draw=black, fill=red!30]
\tikzstyle{io} = [trapezium, trapezium left angle=70, trapezium right angle=110, minimum width=3cm, minimum height=1cm, text centered, draw=black, fill=blue!30]
\tikzstyle{process} = [rectangle, minimum width=3cm, minimum height=1cm, text centered, text width = 4cm, draw=black, fill=orange!30]
\tikzstyle{decision} = [diamond, minimum width=3cm, minimum height=1cm, text centered, text width = 4cm, draw=black, fill=green!30]
\tikzstyle{arrow} = [thick,->,>=stealth]

\usepackage{graphicx}

\usepackage[normalem]{ulem}
\usepackage{appendix}
\usepackage{color}

\usepackage{pdfpages}
\usepackage{chngpage}
\def\den0{$\rho_0$}

\def\es0{$E_{\rm sym}(\rho_0)$}

\def\us0{$U_{\rm sym}(\rho_0,k_F)$~}

\def\l0{$L(\rho_0)$~}

\newcommand{\beq}{\begin{equation}}
\newcommand{\eeq}{\end{equation}}
\newcommand{\ba}{\begin{array}}
\newcommand{\ea}{\end{array}}
\newcommand{\bea}{\begin{eqnarray}}
\newcommand{\eea}{\end{eqnarray}}
\newcommand{\bi}{\begin{itemize}}  %\setlength{\itemsep}{0\parsep}}
\newcommand{\ei}{\end{itemize}}
\newcommand{\ben}{\begin{enumerate}} %\setlength{\itemsep}{0\parsep}}
\newcommand{\een}{\end{enumerate}}
\newcommand{\bc}{\begin{center}}
\newcommand{\ec}{\end{center}}

\usepackage{pgf}

\usepackage{array}

\usepackage{soul,xcolor}
\setstcolor{blue}
\usepackage{makecell}

\begin{document}

\title{Symbolic Regression for Interpretable Emulation of Proton Collective Flow\\ in Intermediate-Energy Heavy-Ion Collisions} 

\author{Nicholas Cox\footnote{ncox6@leomail.tamuc.edu}}
\affiliation{Department of Physics and Astronomy, East Texas A$\&$M University, Commerce, TX 75429, USA}
\author{Xavier Grundler\footnote{xgrundler@leomail.tamuc.edu}}
\affiliation{Department of Physics and Astronomy, East Texas A$\&$M University, Commerce, TX 75429, USA}
\author{Bao-An Li\footnote{Corresponding author: Bao-An.Li@etamu.edu}}
\affiliation{Department of Physics and Astronomy, East Texas A$\&$M University, Commerce, TX 75429, USA}
\date{\today}

\begin{abstract}
Symbolic regression provides an interpretable machine-learning approach for constructing explicit analytic relations between physical inputs and observables. In this work, we develop symbolic-regression emulators for the isospin-dependent Boltzmann-Uehling-Uhlenbeck (IBUU) transport model and compare their performance with deep neural network (DNN) emulators. Using the same transport-model data employed in our previous emulator studies, we show that symbolic regression can reproduce the proton mid-rapidity slope $F_1$ of transverse flow $v_1$ and elliptic flow $v_2$ with accuracy comparable to that of DNNs, while providing explicit analytic expressions and substantially faster prediction once trained. We further demonstrate the use of symbolic regression in the reverse direction by constructing analytic relations that predict the in-medium nucleon-nucleon cross-section modification factor $X$ from the flow observables. Although the symbolic-regression models require substantially longer training times and exhibit greater run-to-run variation than DNNs, their analytic form and rapid evaluation make them promising tools for future transport-model sensitivity and uncertainty analyses.
\end{abstract}

% \begin{description}
% \item[Background:] 
% \item[Purpose:] 
% \item[Method:] 
% \item[Results:] 
% \item[Conclusions:] 
% \end{description}
% \end{abstract}

\maketitle

\section{Introduction}\label{intro}
In recent years, various types of machine learning have been introduced into nuclear physics as tools for discovering new connections within datasets. Some of these approaches, such as deep neural networks (DNNs) and Gaussian processes (GPs), have been used as emulators to accelerate the generation of datasets for Bayesian analyses that would otherwise require much longer computational times with full numerical simulators \cite{Boehnlein:2021eym,Zhou:2023pti}. 
Specifically, in low- and intermediate-energy heavy-ion collisions, machine
learning has been applied to a range of problems including heavy-ion
collision event characterization, inference of nuclear-matter properties, and
transport-model emulation; see, e.g., Refs.~\cite{He:2023urp,Wang:2023kcg} for recent reviews
and representative applications. 
In particular, neural networks have been used to reconstruct collision
impact parameters~\cite{Bass:1993vx,Li:2020xzh,Li:2021impact,Zhang:2022impact,tsbw-r6tj},
infer neutron-skin characteristics~\cite{Huang:2022neutronskin}, and determine
the deformation and orientation of colliding nuclei
\cite{Yang:2024orientation,Yang:2026orientation} and the temperature reached
in the collision~\cite{Song:2021vkg} from final-state observables. Machine
learning has also been used to identify signatures of the nuclear liquid-gas
phase transition~\cite{Wang:2020qpj} and nuclear symmetry energy
\cite{Wang:2021symmetryML,Wang:2022decoding}, as well as to emulate flow and
stopping observables from transport-model simulations for Bayesian inference
\cite{Li:2023ydi,Li:2025iqq,Cox:2024mkz,Wei:2025ml,Han:2026acn}. While these studies demonstrate the growing
utility of machine learning for intermediate-energy heavy-ion physics, the
use of symbolic regression to obtain explicit analytic representations of
transport-model responses and to explore their inverse mappings remains
largely unexplored. Here we investigate this approach using the
isospin-dependent Boltzmann-Uehling-Uhlenbeck (IBUU) transport model
\cite{Li:1997px,Li:2003ts,Li:2003zg,Li:2008gp}, comparing symbolic-regression and DNN
emulators for both forward and inverse problems.

Symbolic regression is a machine-learning approach that seeks to identify explicit analytic expressions describing relationships between input and output data \cite{Schmidt2009,Udrescu2020}. In contrast to conventional machine-learning models whose learned relationships are generally encoded in large sets of model parameters, symbolic regression produces explicit mathematical expressions that can be inspected and interpreted directly. This provides several potential advantages for physics applications: the resulting expressions can reveal nonlinear correlations and functional dependencies between physical quantities, can be readily evaluated, differentiated, or otherwise manipulated analytically, and can be directly incorporated into sensitivity studies and Bayesian analyses. Moreover, once an analytic expression has been obtained, it can be evaluated repeatedly without retraining the machine-learning model, potentially providing substantial computational savings when millions of model evaluations are required.

Fast surrogate models, or emulators, have become an important component
of Bayesian analyses of computationally expensive nuclear models
\cite{Pratt:2015zsa,Bass:2017determination,Phillips:2021rsc,Drischler:2023wqf}.
In heavy-ion physics, GP emulators have enabled Bayesian
inference by replacing computationally expensive model calculations with
fast surrogate evaluations \cite{Pratt:2015zsa,Bass:2017determination}.
In our previous work, we developed GP and DNN emulators for
the IBUU transport model at intermediate energies
\cite{Li:2023ydi,Li:2025iqq,Cox:2024mkz}.
Here we investigate whether symbolic regression can provide an alternative approach that combines
competitive predictive accuracy with explicit analytic mappings between
model parameters and observables.

The rest of the paper is organized as follows. Section~\ref{mm} describes
the models and methods used in this work, while Section~\ref{res} presents
the emulator and inverter results, including the effects of different
training and testing data splits and the speed tests. Section~\ref{sum}
summarizes our main findings. To avoid clutter, the selected equations and
the speed-test results are collected in Appendices~\ref{chosenEq} and
\ref{speed}, respectively.

\section{Models and Methods}\label{mm}
The present study uses the IBUU simulations and datasets previously
generated for Bayesian inference of transport properties and the stiffness
of nuclear matter \cite{Li:2023ydi,Li:2025iqq} from proton collective-flow data taken by the FOPI and HADES Collaborations \cite{Reisdorf2012,AdamczewskiMusch2023}.
In these Bayesian analyses, two input model parameters are varied. The first is the ratio of the nucleon-nucleon (NN) in-medium and free-space cross sections,
$X \equiv \sigma_{NN}^{\rm med}/\sigma_{NN}^{\rm free}$.
The second is the incompressibility of the symmetric nuclear matter (SNM) equation of state $E_0(\rho)$, defined as
$K=9\rho_0^2[d^2E_0(\rho)/d\rho^2]$ at the SNM saturation density $\rho_0=0.16$~fm$^{-3}$. Of the simulation results, we focused on the transverse flow at mid-rapidity in the center-of-mass frame, $v_1$, specifically its mid-rapidity slope, $F_1$, which is more commonly used, and the elliptic flow, $v_2$. These observables have been measured at several beam energies at GSI by the FOPI \cite{Reisdorf2012} and HADES \cite{AdamczewskiMusch2023} collaborations.

Throughout, we will refer to a model which takes $F_1$ and $v_2$ as inputs to predict $X$ or $K$ as an \emph{inverter}. A model which instead takes $X$ and $K$ as inputs to predict $F_1$ or $v_2$ will be called an \emph{emulator}. This distinction is useful because the two directions of the mapping address different scientific questions: the emulator approximates the forward transport-model simulation, whereas the inverter attempts to extract information about the underlying nuclear-matter properties from experimentally accessible observables. 
Such explicit inverse mappings could in future be used to propagate
experimental uncertainties in $F_1$ and $v_2$ to uncertainties and
correlations in the inferred $X$ and $K$, and to examine nonlinear
biases associated with finite measurement precision, including
Jensen-type effects \cite{Jen,Ber,Ebook,infobook}. Such applications could provide a transparent framework for quantifying how uncertainties in measured collective flow propagate back to the underlying nuclear-matter properties, similar to such studies \cite{Li:2026ult,Li:2026NS} in Bayesian inferences of EOS parameters from observational data of neutron stars.

\subsection{DNN model for Comparison}
The DNN model used for comparison is very similar to the Scikit-Learn neural network shown in Fig. 2 of Ref. \cite{Cox:2024mkz}, but instead of using a single neural network for both output parameters, two neural networks with one output each are used. This configuration matches the symbolic-regression models and enables a consistent comparison between the two methods. It also allows each DNN to be optimized specifically for its corresponding output dataset.

Another change from Ref.~\cite{Cox:2024mkz} is that the maximum number of iterations of the L-BFGS regressor was increased from 80 to 400. This change allowed the models to converge, while only slightly affecting their accuracy, suggesting that the additional iterations do not lead to significant overfitting. As in Ref.~\cite{Cox:2024mkz}, the data were scaled using the Scikit-Learn StandardScaler, with 75\% of the data used for training and 25\% for testing. The loss function for the DNN was the mean squared error (MSE) between the DNN output and the target value,
\begin{equation}
\mathrm{MSE} = \frac{1}{n}\sum_{i=1}^{n}
\left(X_{\mathrm{pred},i}-X_{\mathrm{target},i}\right)^2 .
\end{equation}
This is the default loss function used by the Scikit-Learn implementation and is appropriate for the present regression problem.

\subsection{PySR Symbolic Regression Package and Model}

The symbolic-regression calculations were performed using the PySR package,
which provides a Python interface to the Julia-based
\texttt{SymbolicRegression.jl} package \cite{Cranmer:2023}. PySR allows
the user to control the mathematical operators, loss function, expression
complexity, and other constraints that define the search space. These
features are particularly useful here because we seek relatively simple
analytic expressions that can be compared directly with the DNN models.

To facilitate a direct comparison with the DNN models, we used the MSE as the loss function. 
The maximum expression complexity was set to 20, and the number of PySR iterations was set to 400.
In PySR, expression complexity is determined by the number of nodes in an expression rather than solely by
its polynomial order. For example, the expressions $2\cdot x+3$ and
$x^2/y$ have comparable complexity because constants, variables, operators,
and powers each contribute to the node count. Consequently, a higher-order
polynomial will generally have greater complexity, although the relationship
between polynomial order and complexity is not one-to-one.

To further improve the interpretability of the resulting expressions, we
restricted the symbolic-regression search to rational functions. The allowed
binary operators were addition, subtraction, multiplication, and division,
while the allowed unary operators were the second, third, and fourth powers.
We further imposed nesting constraints on these unary operators, as
summarized in \autoref{tab:NestConstraint}. These constraints limit the
maximum order of nested powers and thereby restrict the complexity of the
generated rational functions. In particular, the resulting expressions
cannot contain a nested power higher than eighth order. They also reduce the
number of possible pathways through which higher-order expressions can be
generated, thereby favoring simpler functional forms.

\begin{table}[ht]
    \centering
    \caption{Nesting constraints for the PySR model}
    \label{tab:NestConstraint}
    \begin{tabular*}{\columnwidth}{@{\extracolsep{\fill}} c c c}
        \hline
        \makecell{External \\ Operator Type} & \makecell{Allowed\\Nested Operators} & Example\\\hline
        Quartic & None & (X)$^4$  \\
        Cubic & 1 quadratic & (3X$^2$ - 8)$^3$ \\
        Quadratic & \makecell{1 quadratic, cubic, \\ and quartic} & (X$^4$+Y$^3$-X$^2$)$^2$\\
        \hline
    \end{tabular*}
\end{table}

The symbolic-regression search proceeds through successive generation,
modification, simplification, and optimization of candidate expressions.
The evolutionary search modifies the structure of candidate expressions,
including their number and arrangement of nodes. Candidate expressions are
then simplified using algebraic simplification procedures, while their
numerical coefficients are optimized to minimize the loss on the training
data. Unlike the DNN calculations, the symbolic-regression calculations
reported in this work were performed using the original, unnormalized
input and output variables. We nevertheless tested the effect of
normalization separately, as discussed in Sec.~\ref{nor}.

Because the search is stochastic and is performed for a finite number of
iterations, different runs do not necessarily produce the same analytic
expression. Multiple expressions can provide comparably accurate
descriptions of the same input-output relationship. We therefore performed
multiple independent runs and developed criteria for selecting representative
expressions based on their accuracy, complexity, absence of singularities
within the relevant data domain, and reproducibility. These criteria are
described in the following subsection.

\subsection{How Equations Were Chosen}

The equations reported in the tables in the Appendix A were selected from 25 independent
runs of the PySR symbolic-regression model. For each run, we recorded both
the equation with the lowest MSE and the equation identified by PySR as the
``best'' equation. The latter is selected by balancing accuracy against
expression complexity. Specifically, PySR considers the most accurate
equation at each complexity level and evaluates the improvement in the loss
relative to the corresponding increase in complexity. This procedure
effectively identifies equations along the Pareto frontier, which represents
the trade-off between model accuracy and expression complexity. Equations on
this frontier provide increasingly accurate descriptions of the data while
requiring increasingly complex functional forms.

PySR assigns a score to equations along the Pareto frontier according to
\begin{equation}
\mathrm{Score}
=
-\frac{\Delta\log(\mathrm{MSE})}
{\Delta\mathrm{Complexity}},
\end{equation}
with a larger score corresponding to a greater reduction in the loss per
unit increase in complexity. The equation with the largest score is
identified by PySR as the ``best'' equation. In a small number of cases,
however, we selected an equation based primarily on its accuracy rather than
the PySR score, as discussed below.

In addition to accuracy and complexity, we required the selected equations
to be well behaved throughout the relevant domain of the dataset. In
particular, equations containing singularities within the sampled domain
were excluded, since such singularities can lead to unphysical or unstable
predictions in regions where the model is expected to be applicable.

\begin{figure*}
    \centering
    \includegraphics[width=0.95\linewidth]{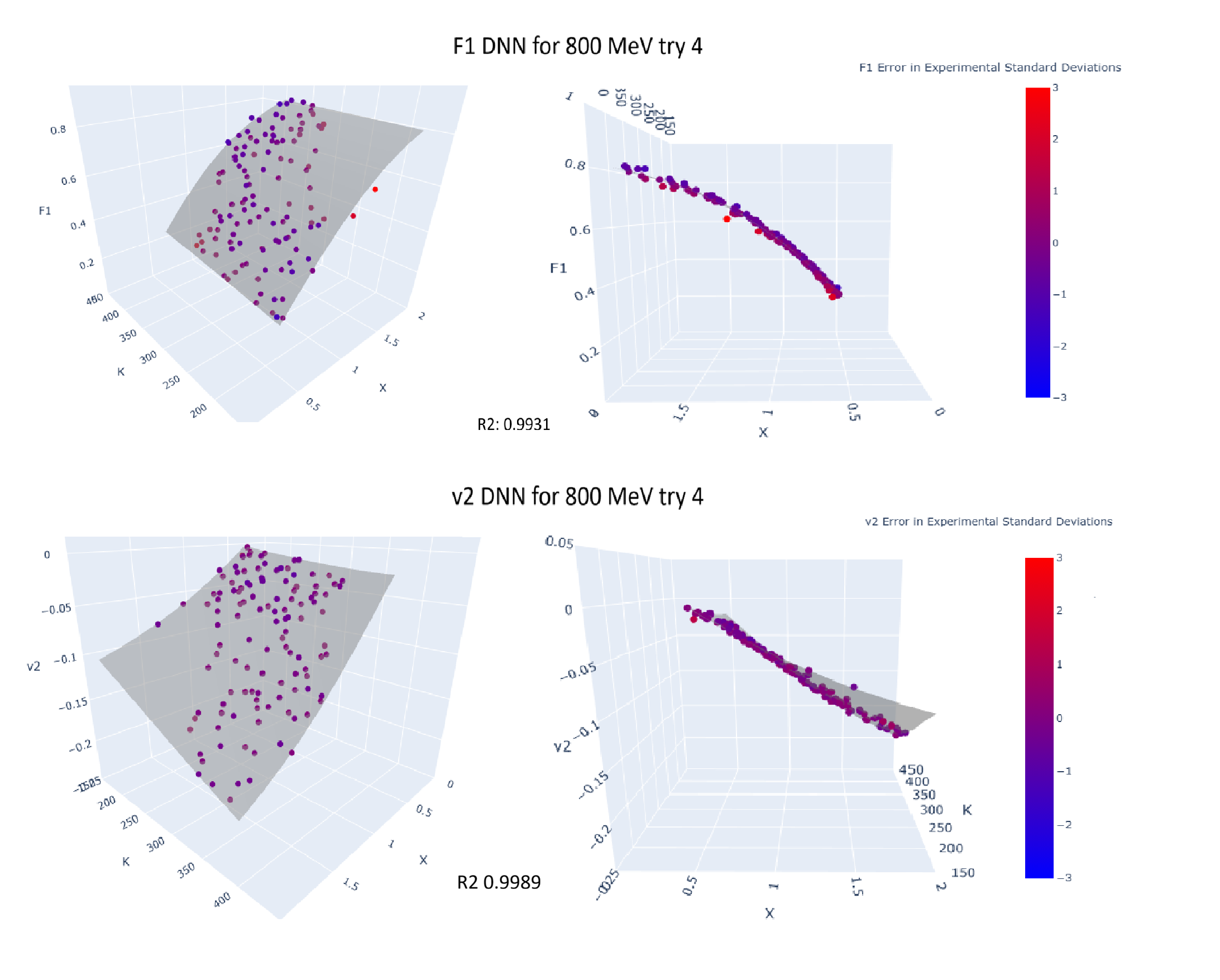}
    \caption{Examples of surfaces of F1 and v2 with respect to X and K created by a DNN for mid-central Au+Au reactions at a beam energy of 800 MeV/nucleon. There are similar static and interactive surfaces available on GitHub \cite{Nic-data2} for the rest of the beam energies listed in Tables~\ref{XEqu},
\ref{KEqu}, \ref{F1Equ}, and \ref{v2Equ}. The symbols are results of IBUU simulations for these reactions \cite{Li:2023ydi,Li:2025iqq}.}
    \label{fig:DNNsurfaces}
\end{figure*}

\begin{figure*}
    \centering
    \includegraphics[width = 0.95\linewidth]{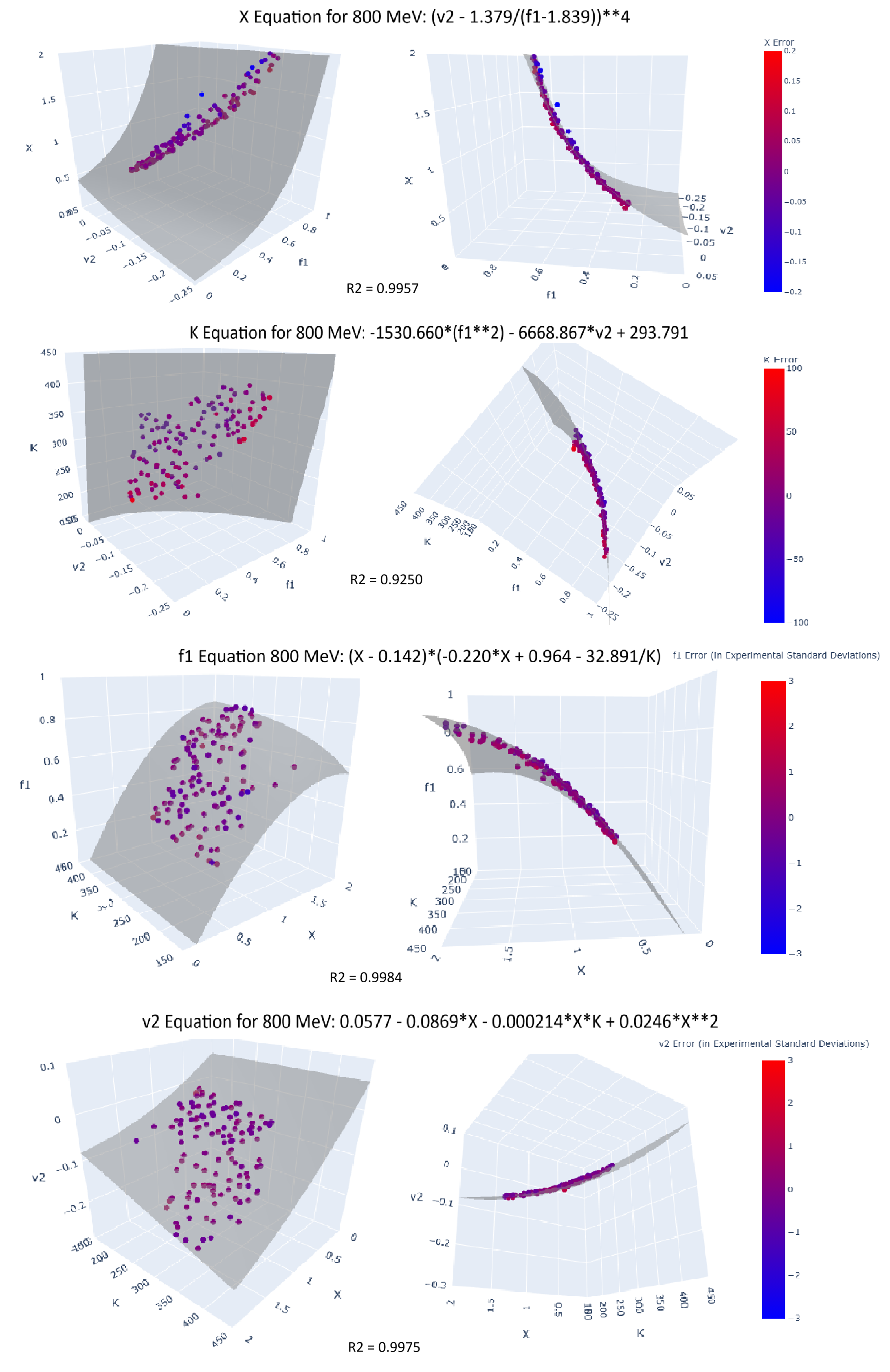}
    \caption{Surfaces generated by the symbolic regression for all of the variables at 800 MeV. Interactive 3D plots are available at \cite{Nic-data2}.}
    \label{fig:Symregsurfaces}
\end{figure*}

The overall predictive accuracy of the candidate equations was assessed
using the coefficient of determination,
\begin{equation}
R^2 = 1-\frac{\mathrm{SSE}}{\mathrm{TSS}},
\end{equation}
where SSE and TSS denote the sum of squared errors and the total sum of
squares, respectively. When candidate equations had similar $R^2$ values,
we used their complexity as an additional selection criterion, favoring the
simpler expression. 
When multiple equations had comparable accuracy and complexity but
produced noticeably different three-dimensional surfaces, we used the
corresponding DNN surface only as a tie-breaking criterion and selected
the equation whose surface most closely resembled the DNN result. This criterion was motivated by the generally
higher accuracy and greater run-to-run consistency of the DNN models. Examples
of the DNN-generated surfaces are shown in \autoref{fig:DNNsurfaces}, while
the corresponding surfaces from symbolic regression are shown in
\autoref{fig:Symregsurfaces}. Interactive versions of these plots are
available at \cite{Nic-data2} and were generated using the Plotly Python
package \cite{plotly2026}.

Finally, when several candidate equations satisfied the above criteria
similarly well, preference was given to expressions that were reproduced more
consistently across independent PySR runs. This additional criterion favors
functional relationships that are less sensitive to the stochastic nature of
the symbolic-regression search and therefore are more likely to be
reproducible.

\subsection{Testing Model Speed}

To compare the computational speed of the DNN and symbolic-regression emulators, we incorporated each emulator into the Bayesian framework using the Metropolis--Hastings approach to Markov Chain Monte Carlo (MCMC) sampling as in Refs.~\cite{Li:2023ydi,Li:2025iqq}. In the present test, however, the sampling framework
was used solely as a standardized procedure for evaluating emulator
performance. Each emulator was called $10^6$ times to calculate $F_1$ and
$v_2$ from randomly generated pairs of $X$ and $K$ distributed over a
two-dimensional Latin hypercube covering their respective prior ranges
\cite{Li:2023ydi,Li:2025iqq}. 

For each model, we recorded both the training time and the total time
required for the $10^6$ emulator evaluations. The training time provides a
measure of the computational cost of constructing the emulator, while the
evaluation time characterizes the cost of repeatedly using the trained
model. We also recorded the number of accepted MCMC steps to verify
that differences in the measured execution times were not caused by
different numbers of steps being written to the output files.

\section{Results and Discussion}\label{res}

Our main symbolic-regression results are summarized in Tables~\ref{XEqu},
\ref{KEqu}, \ref{F1Equ}, and \ref{v2Equ} in Appendix~\ref{chosenEq}, while the results
of the speed tests are given in Tables~\ref{DNNTT} and
\ref{SymRegTT} in Appendix~\ref{speed}. In addition to the static figures presented
below, interactive versions of the three-dimensional plots are available
in Ref.~\cite{Nic-data2}, allowing the surfaces generated by the models to
be viewed from different directions.

\subsection{Emulator}

Both the DNN and symbolic-regression models provide accurate descriptions
of $F_1$ and $v_2$ as functions of $X$ and $K$. For all FOPI datasets
simulated with IBUU \cite{Li:2025iqq}, both approaches consistently achieve
an average $R^2$ greater than 0.98. For the HADES dataset
\cite{Li:2023ydi}, the DNN achieves an average $R^2$ of 0.97 with little
run-to-run variation, whereas the symbolic-regression models give
$R^2$ values ranging from 0.80 to 0.99 across five independent runs. These
results are obtained using the 25\% testing-data subset for both methods.
Overall, the symbolic-regression emulator exhibits greater run-to-run
variation in $R^2$ than the DNN emulator for all datasets.

\begin{figure*}
    \centering
    %\vspace{-50pt}
    \begin{tabular}{cc}\includegraphics[width=0.95\linewidth]{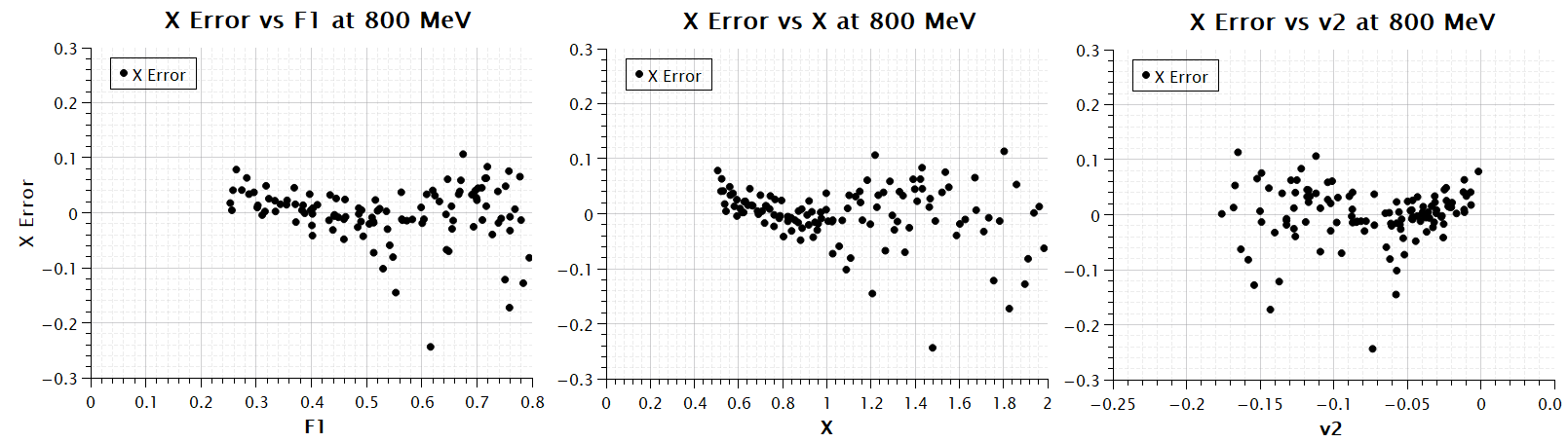} \\
    \includegraphics[width=0.95\linewidth]{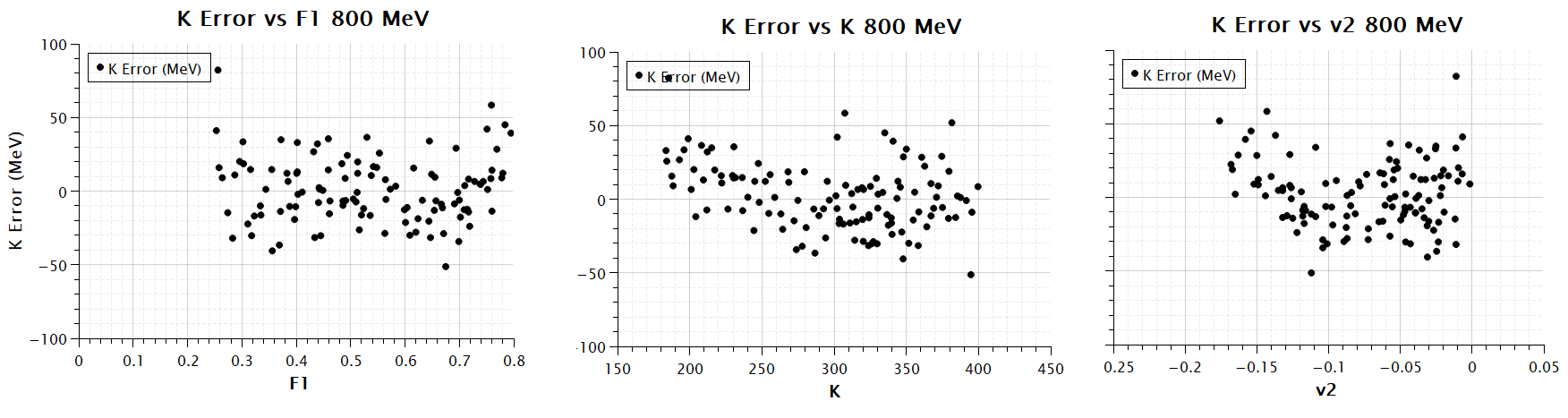} \\
    \includegraphics[width=0.95\linewidth]{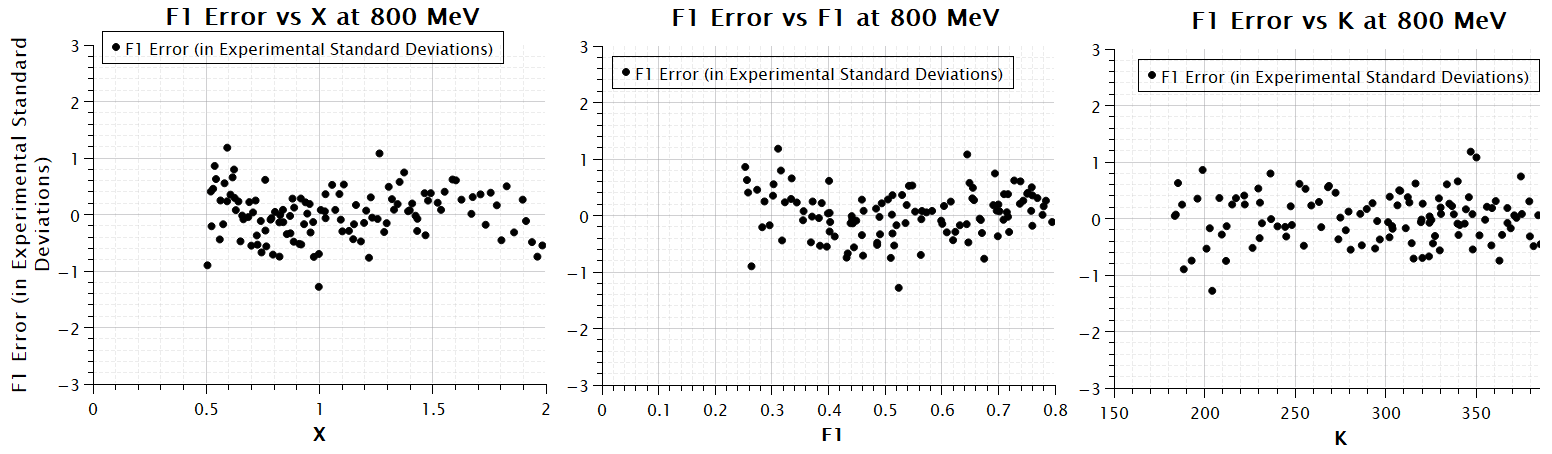} \\
    \includegraphics[width=0.95\linewidth]{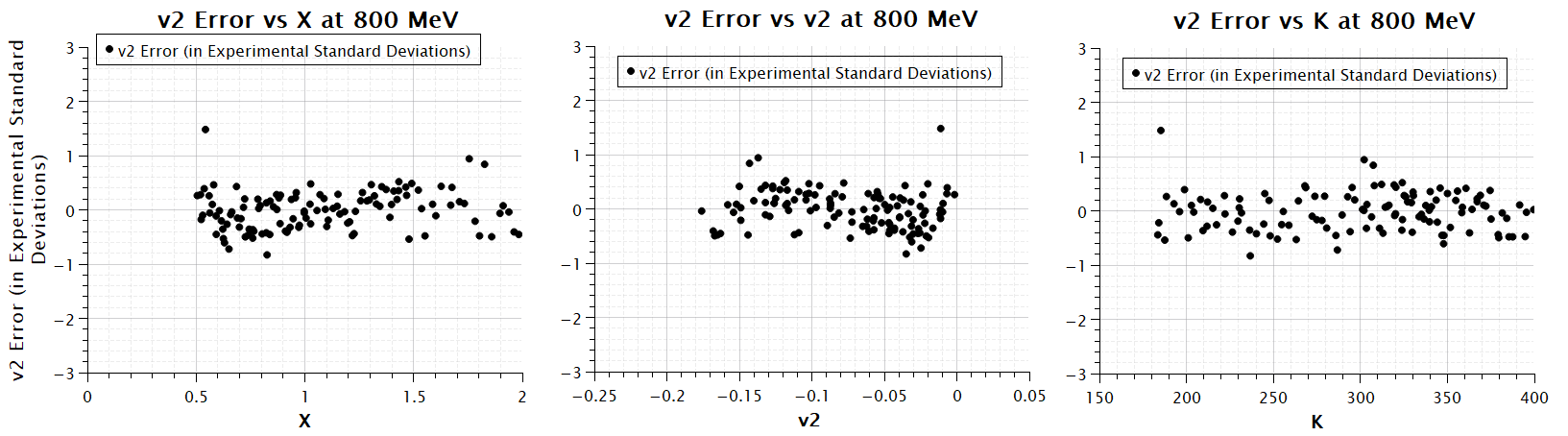}
    \end{tabular}
    \caption{Residuals generated by the functions vs the actual data as a function of each variable, this corresponds to the functions shown in \autoref{fig:Symregsurfaces}.}
    \label{fig:SymregResiduals}
\end{figure*}

The larger variation in symbolic-regression performance is a consequence
of the stochastic search over a large space of possible analytic
expressions. Multiple expressions with different structures can provide
similarly accurate descriptions of the same input-output relationship,
leading to different $R^2$ values in different runs. In contrast, the DNN generally produces similar predictions for a
given dataset, resulting in more consistent $R^2$ values between runs.
Despite this greater variation, symbolic regression does not substantially
degrade the predictive accuracy relative to the DNN emulator. The analytic
expressions generated by symbolic regression also provide a better
description of the output data than the linear regressions used in our
previous work \cite{Cox:2024mkz}. 
The residuals show little systematic dependence on either the input or
output variables, suggesting that no obvious additional functional
dependence is required to describe the remaining deviations within the
accuracy of the present datasets. Examples of the residuals used in this
analysis are shown in \autoref{fig:SymregResiduals}. For a given dataset,
the emulators generally provide more accurate predictions in regions of
moderate $F_1$ and $v_2$ than near their extreme values. This behavior is
likely related to the greater density of training points in the central
regions of the data, where the models are consequently more strongly
constrained.

The selected symbolic-regression expressions for $F_1$ and $v_2$ for all
datasets are given in \autoref{F1Equ} and \autoref{v2Equ}, respectively.

\subsection{Inverter}

Both the DNN and symbolic-regression inverters were able to predict $X$
from $F_1$ and $v_2$ with good accuracy for the FOPI datasets. 
For clarity, we note that the values quoted below refer to the performance averaged over five
independent runs, whereas the equations listed in Table \ref{XEqu} correspond to
the selected representative expressions. For all FOPI beam energies, both approaches achieved $R^2>0.9$. As for the
emulators, 25\% of the data were reserved for testing. Across five
independent runs, the variation in $R^2$ for the DNN inverter was less than
0.02, whereas the symbolic-regression inverter exhibited variations larger
than 0.05 for some datasets. This behavior is consistent with that observed
for the emulators and reflects the stochastic search over different
candidate analytic expressions. Despite the larger run-to-run variation,
the average $R^2$ values at each beam energy were comparable for the two
inverters.

Both inverters perform less well for the HADES reaction, for which we have 90 simulated datasets, than for the FOPI reactions, which have 120 at each beam energy. This difference may partly reflect the smaller HADES training set, although differences in the underlying parameter dependence of the two datasets may also contribute. The DNN inverter provided only a modest improvement over the linear regression used in Ref.~\cite{Cox:2024mkz}, while the symbolic-regression inverter performed
worse than the linear regression in four of the five runs. In the selected
``best'' symbolic-regression expression, $v_2$ does not appear explicitly.
This behavior is plausible because $F_1$ and $v_2$ are strongly
anticorrelated in these datasets \cite{Li:2023ydi,Li:2025iqq}. Once the
dominant dependence of $X$ on $F_1$ has been captured, additional $v_2$
terms may provide insufficient reduction in the loss to compensate for
their increase in expression complexity. When equations were selected
solely on the basis of their minimum MSE, some expressions achieved a
noticeable improvement in $R^2$ but contained singularities within the
relevant data domain. Such expressions were therefore excluded from
consideration. As for the emulators, the residuals show no obvious
systematic dependence on the input variables, suggesting that the remaining
prediction errors are primarily associated with the scatter in the data.
The selected expressions for $X$ are given in \autoref{XEqu}.

The prediction of $K$ is considerably more challenging for both inverters,
particularly at lower beam energies. At 150 MeV/nucleon, for example, the
average $R^2$ is only 0.35. The performance improves with increasing beam
energy, reaching a maximum average $R^2$ of 0.89 at 800 MeV/nucleon, before
decreasing somewhat at the highest beam energies. This behavior is
consistent with the broad posterior distributions of $K$ found in our
previous Bayesian analyses, which also exhibit beam-energy-dependent
substructure \cite{Li:2023ydi,Li:2025iqq}. The DNN consistently outperforms
the symbolic-regression inverter in predicting $K$ from the flow
observables. It also exhibits substantially better run-to-run consistency.
Across five independent runs, the $R^2$ of the symbolic-regression inverter
varies by more than 0.1 at several beam energies, indicating that the
resulting analytic expressions for $K$ are less reproducible than those for
$X$.

For the HADES dataset, the DNN inverter improves substantially upon the
linear-regression result of Ref.~\cite{Cox:2024mkz}, increasing $R^2$ by
more than 0.2 in four of the five runs. In contrast, the symbolic-regression
inverter improves upon the linear model for $K$ in only one of the five
runs. The selected expressions for $K$ are given in \autoref{KEqu}.

Several factors may contribute to the poorer performance of symbolic
regression in predicting $K$. One possibility is that the dependence of
$K$ on the flow observables is not well represented by the restricted
rational-function basis adopted in this work. This possibility is suggested
by the three-dimensional surfaces generated by the DNNs. Some of the more
accurate DNN surfaces exhibit structures that would be difficult to represent
with low-order rational functions without introducing singularities. For
example, a few surfaces display features resembling oscillatory or
``ripple-like'' behavior that could potentially be represented more
naturally by functions such as sine or cosine, which were excluded from the
present symbolic-regression search.

\begin{figure}
    \centering
    \includegraphics[width=\linewidth]{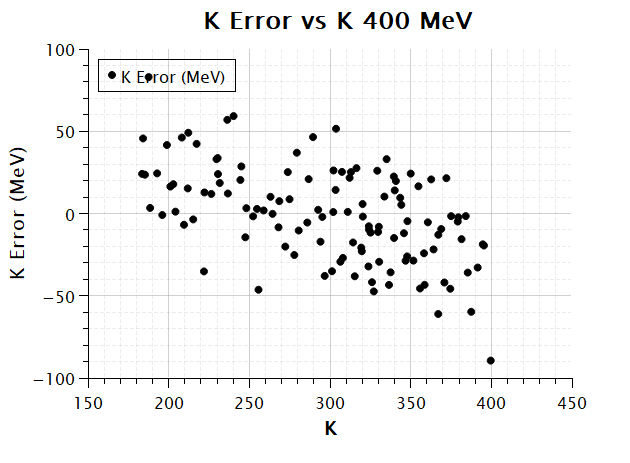} \\
    \caption{Example of a downward trend of errors against the K value.}
    \label{fig:KResEx}
\end{figure}

Another possibility is that the two flow observables do not contain
sufficient information to constrain $K$ accurately over the full parameter
range. The residuals provide some evidence in this direction. Although they
show no clear systematic dependence on $F_1$ or $v_2$, they exhibit a
consistent decrease with increasing $K$ at all beam energies. Thus, the
inverters tend to overpredict $K$ at low values and underpredict it at high
values, as illustrated in \autoref{fig:KResEx}. Such a trend indicates a systematic component in the prediction error that
is not captured by purely random residual fluctuations.

The residual trend could arise from the broad spread of $K$ relative to the
flow observables, making it intrinsically more difficult for the available
data to constrain $K$. It could also indicate that additional physical
variables or observables are needed to describe the dependence of $K$ more
completely. Alternatively, the trend could result from limitations of the
restricted functional form used in the symbolic-regression search. The
present results do not distinguish between these possibilities, but they
clearly demonstrate that predicting $K$ is more challenging than predicting
$X$ from the two flow observables considered here. 

It is worth emphasizing that the present study uses only two observables,
the proton mid-rapidity slope of the directed flow, $F_1$, and the elliptic
flow, $v_2$, to constrain the two parameters $X$ and $K$. Thus, for each
reaction, the analysis provides a deliberately simple two-observable
example of the broader Bayesian inference problem. The experimental data
from HADES and FOPI, however, contain substantially more information,
including the rapidity and transverse-momentum dependence of proton and
cluster flows, as well as other observables such as particle yields and
their rapidity dependence, which provide information on nuclear stopping.
A simultaneous analysis of these observables would be highly desirable but
would require significantly more transport-model simulations and analyses. 

In the future, the framework could therefore be extended from the present
two-observable--two-parameter example to the more realistic problem of
constraining a small number of transport and nuclear-matter parameters
using many experimental observables. Such an extension could help break
parameter degeneracies and provide stronger and more robust constraints.
At the same time, the corresponding inverse problem would be more
challenging: an inverter for $X$ or $K$ would generally involve multiple,
potentially correlated observables as inputs, and a unique analytic
mapping need not exist. The increased dimensionality of the symbolic
search would also provide an additional challenge. These issues make the
development of fast, forward emulators particularly valuable, since they
allow the full many-observable likelihood to be explored without requiring
a separate analytic inversion for each parameter. The present
two-observable--two-parameter study thus provides a controlled starting
point for this broader program.

\subsection{Speed Test}

To compare the computational speed of the DNN and symbolic-regression
emulators within the Bayesian-analysis framework, we recorded the training
time and the time required to complete the analysis, which required one million model evaluations.
We also compared the number of accepted MCMC steps for the two
models to ensure the associated file-writing time did not affect the measured prediction
times.

The DNN emulator required a much shorter training time, approximately
0.1~s, whereas training the symbolic-regression emulator required up to
several minutes. This difference reflects the substantially more
computationally intensive stochastic search over candidate analytic
expressions required by symbolic regression. Because of the consistency
issues discussed earlier, multiple symbolic-regression runs may also be
required to identify a satisfactory and reproducible expression, further
increasing the effective training cost.

Once trained, however, the symbolic-regression emulator was substantially
faster to evaluate. When each emulator was called $10^6$ times, the
symbolic-regression model required only a few seconds, whereas the DNN
required several minutes. Thus, although symbolic regression is
substantially more expensive to train, its much lower evaluation cost
makes it advantageous when a trained emulator is to be called a very
large number of times. The two models produced very similar numbers of
accepted MCMC steps at each beam energy, with the symbolic-regression
model having marginally more accepted steps in most cases. Thus, differences in
the number of accepted steps and the associated file-writing time cannot
account for the large difference in prediction time. 
Interestingly, direct evaluation of the stored analytic expressions,
without invoking the PySR model infrastructure, further eliminates the
training cost and retains a prediction time of only tens of seconds for
$10^6$ evaluations. The complete speed measurements are given in \autoref{DNNTT} for the DNN emulator and \autoref{SymRegTT} for the symbolic-regression emulator.

\subsection{Effects of Testing Data Split Size and Normalization}\label{nor}

We also investigated the sensitivity of the symbolic-regression emulator to
the choice of training and testing data sizes. The training and testing subsets were randomly selected from the simulated datasets, so that both subsets sample the same ranges of $X$ and $K$.
The fractions of the data reserved for testing were 5\%, 10\%, 15\%, 25\%, and 50\%. For each testing
fraction, the symbolic-regression model was able to obtain accurate
expressions for both $F_1$ and $v_2$, regardless of whether the ``best''
or ``accuracy'' selection criterion was used. To assess the dependence of
the emulator accuracy on the testing-data fraction, we performed five
independent runs for each choice. The average $R^2$ values showed no
significant dependence on the testing fraction for either $F_1$ or $v_2$.
The 5\% and 50\% testing fractions generally gave $R^2$ values about 0.01
lower than the intermediate 10\%, 15\%, and 25\% fractions. These small
differences may reflect the competing effects of a smaller training set for
the 50\% split and a smaller testing set for the 5\% split. Overall, the
results indicate that the emulator is relatively insensitive to the choice
of training/testing split over the range considered here, and all of the
tested splits provide adequate accuracy for $F_1$ and $v_2$.

For the inverse problem of predicting $X$ and $K$ from $F_1$ and $v_2$, the
10\%, 15\%, and 25\% testing fractions tended to give the best average
performance. However, the run-to-run variations are substantially larger
for the inverter than for the emulator, making this trend less conclusive.
Even after averaging over five independent runs, the $R^2$ values for the
predicted $X$ varied by up to 0.05 among the different testing fractions,
while the corresponding variation for $K$ was as large as 0.5, without a
clear systematic dependence on the testing fraction. Thus, within the
statistics of the present study, the choice of testing fraction appears to
have a relatively small effect on the emulator, whereas its effect on the
inverter is more difficult to establish. A larger number of independent
runs and/or a larger training dataset would be needed to determine this
dependence more definitively.

We also investigated the effect of normalizing the input and output
variables before performing symbolic regression. The resulting $R^2$
values were comparable to those obtained using the unnormalized data,
indicating that the symbolic-regression procedure can accommodate the
different numerical scales of the variables considered here. In
particular, $X$ varies from approximately 0 to 2, while $K$ ranges from
about 150 to 450~MeV. We nevertheless use the unnormalized data throughout
this work for two reasons. First, the resulting analytic expressions retain the original physical
scales and units of the input and output variables.
Second, some of the expressions obtained from normalized data developed
singularities when transformed back to the original variables. Using
unnormalized data therefore avoids these problematic expressions while
preserving comparable predictive accuracy.

\section{Summary}\label{sum}
In summary, we have developed and tested symbolic-regression emulators for
proton directed and elliptic flow in intermediate-energy heavy-ion
collisions simulated with the IBUU transport model and compared them with
DNN emulators using the same datasets. For the forward problem, in which
$X$ and $K$ predict the proton mid-rapidity flow observables $F_1$ and
$v_2$, symbolic regression achieves accuracy comparable to the DNN. For
the FOPI datasets, both approaches generally achieve $R^2>0.98$, while
symbolic regression provides explicit analytic mappings that can be
inspected and analyzed analytically and graphically. 
Although substantially slower to train, the symbolic-regression emulator
is typically one to two orders of magnitude faster than the DNN for
$10^6$ evaluations. Once a satisfactory analytic expression has been
obtained, it can also be reused in subsequent calculations, eliminating
the training step.

For the inverse problem, both DNN and symbolic-regression inverters
predict the in-medium cross-section modification factor $X$ accurately
from the two flow observables for the FOPI datasets, with $R^2>0.9$. The symbolic-regression inverter is less consistent between runs but provides useful analytic relations between $X$,
$F_1$, and $v_2$. In contrast, predicting the incompressibility $K$ is more
challenging, particularly at low beam energies, and the DNN consistently
outperforms symbolic regression. The systematic residual dependence on $K$
suggests that the two flow observables do not fully constrain $K$ and/or
that the restricted rational-function basis does not adequately represent
its dependence.

Although the present inverse analysis uses only $F_1$ and $v_2$, the
framework can ultimately be extended to many observables, for which fast
forward emulators may be more useful than explicit analytic inverse
mappings.

Finally, the emulator accuracy is relatively insensitive to the
training/testing split, while the inverter shows larger variations,
particularly for $K$. Comparable accuracy was obtained with normalized and unnormalized data;
we use the latter to retain the original physical scales and units and
avoid singularities encountered when transforming some normalized
expressions. Overall, symbolic regression provides an attractive complement
to DNN emulators by combining competitive accuracy with explicit analytic
expressions and rapid repeated evaluation, making it promising for future
transport-model sensitivity studies, understanding uncertainty propagation, and Bayesian
analyses requiring large numbers of model evaluations.
\\

\textit{Acknowledgment-}
 This work was supported in part by the U.S. Department of Energy, Office of Science, under Award Number DE-SC0013702, and the NASA-Texas Space Grant Consortium. 
\\

\textit{DATA AVAILABILITY:} The data that support the findings of this article are openly available \cite{Nic-data2}.

\appendix

\section{Tables of Chosen Equations}\label{chosenEq}
In the following tables, we present the chosen equations for $F_1$, $v_2$, $X$ and $K$, respectively. The datasets are labeled using the beam energy per nucleon in units of MeV. The FOPI datasets are for mid-central Au+Au reactions at a beam energy of 150, 250, 400, 600, 800, 1000 and 1200 MeV/nucleon \cite{Reisdorf2012}. The HADES data is for mid-central Au+Au reactions at 1230 MeV/nucleon \cite{AdamczewskiMusch2023}. 

\begin{table}[ht]
    \centering
    \begin{tabular}{|c|c|c|}
        \hline
       Dataset & $F_1$ Equation & $R^2$ \\\hline
        150 & $(0.696-0.182X)(X - \frac{40.614}{K})$ & 0.9948 \\
        250 & $-0.173X^2 + 0.763X + 0.000394K - 0.218$ & 0.9920 \\
        400 & \makecell{$X(-0.204X + 0.000386K + 0.742)$ \\ $- 0.169$} & 0.9925 \\
        600 & $X(0.757-0.163X) - \frac{25.937}{K}$ & 0.9970 \\
        800 & $(X - 0.142)(-0.220X + 0.964 - \frac{32.891}{K})$ & 0.9984 \\
        1000 & $(1.038 - \frac{7.047}{K} - \frac{0.0854}{X})^8$ & 0.9993 \\
        1200 & $(X-0.135)(-0.221X+0.972 - \frac{40.286}{K})$ & 0.9975 \\
        HADES & $X(-0.135X + 0.000603K + 0.342)$ & 0.9648 \\
        \hline
    \end{tabular}
    \caption{Chosen Equations for $F_1$ in terms of X and K.}
    \label{F1Equ}
\end{table}

\begin{table}[ht]
    \centering
    \begin{tabular}{|c|c|c|}
        \hline
        Dataset & $v_2$ Equation & $R^2$ \\\hline
        150 & $0.0652X^2 - 0.275X + \frac{4.776}{K} + 0.0996$ & 0.9986 \\
        250 & $0.0455X^2 - 0.308X + \frac{16.387X}{K} + 0.0969$ & 0.9962 \\
        400 & \makecell{$X(9.728e^{-5}K(X-3.352) - 0.101)$\\$ + 0.075$} & 0.9994 \\
        600 & \makecell{$-8.154e^{-10}K^3 - 0.112X + 0.0588$} & 0.9495 \\
        800 & \makecell{$0.0246X^2 - 0.000214XK - 0.0869X$ \\$ + 0.0577$} & 0.9975 \\
        1000 & \makecell{$0.0237X^2 - 0.000207XK - 0.0786X$ \\$+ 0.0562$} & 0.9993 \\
        1200 & $K(-0.00105 + \frac{0.0027}{X+2.903})$ & 0.9975 \\
        HADES & $0.0100 - F_1v_2(0.000178 - 2.450e^{-5}F_1)$ & 0.9906 \\
        \hline
    \end{tabular}
    \caption{Chosen Equations for $v_2$ in terms of X and K.}
    \label{v2Equ}
\end{table}

\begin{table}[ht]
    \centering
    \begin{tabular}{|c|c|c|}
        \hline
        Dataset & X Equation & $R^2$ \\\hline
        150 & $-F_1^3 + (1.982v_2 - 0.840)^4$ & 0.9945 \\
        250 & $\frac{9.379}{(F_1-2.247)^4}$ & 0.9795 \\
        400 & $2.481(F_1^6 + v_2 +F_1) $ & 0.9837 \\
        600 & $5.355v_2 + 2.481 (F_1^4+F_1)$ & 0.9199 \\
        800 & $(v_2 - \frac{1.379}{F_1-1.839})^4$ & 0.9957 \\
        1000 & $(2.052F_1^2 + 4.095v_2 + 0.627)^2$ & 0.9899 \\
        1200 & $11.330v_2 + (F_1 - v_2 + 0.580)^3$ & 0.9882 \\
        HADES & $2.800F_1$ & 0.9207 \\
        \hline
    \end{tabular}
    \caption{Chosen Equations for X in terms of $F_1$ and $v_2$.}
    \label{XEqu}
\end{table}

\begin{table}[ht]
\vspace{0.5cm}
    \centering
    \begin{tabular}{|c|c|c|}
        \hline
        Dataset & K Equation & $R^2$ \\\hline
        150 & $1254.021F_1 + 2939.542v_2$ & 0.4811 \\
        250 & $\frac{994.757(0.0387-v_2)}{F_1}$ & 0.5295 \\
        400 & \makecell{$-7251.772v_2 - 1789.963(F_1 + 0.168)^2$ \\ $+ 383.370$} & 0.8085 \\
        600 & \makecell{$0.648(29.243F_1 - v_2\frac{F_1-130.718}{F_1+0.294} - 30.113)^2$ \\ $- 204.325$} & 0.8199 \\
        800 & $-1530.660F_1^2 - 6668.867v_2 + 293.791$ & 0.9250 \\
        1000 & $-1470.710F_1^2 - 6924.329v_2 + 316.985$ & 0.9808 \\
        1200 & \makecell{$-7464.236v_2 - 1077.768(-F_1+v_2)^2$ \\$- \frac{F_1}{0.300} + 285.179$} & 0.8584 \\
        HADES & $\frac{29.89 - 2402.832v_2}{F_1}$ & 0.7587 \\
        \hline
    \end{tabular}
    \caption{Chosen Equations for K in terms of $F_1$ and $v_2$.}
    \label{KEqu}
\end{table}

\section{Speed Comparison Tables}\label{speed}
Shown in Table \ref{DNNTT} and Table \ref{SymRegTT}
are time trials for DNN and symbolic regression emulators, respectively, across all FOPI and HADES datasets. The models were called 1 million times across 5 runs, and the times presented here are averages across those runs.

In this set of runs, the increased training time for some beam energies is a result of one run that took excessively long, while the other 4 runs were on the order of 100 seconds. When replacing the symbolic emulator with a previously generated function, the training time is essentially skipped, while the prediction time is larger by a factor of 2-4, taking on average 23-28 seconds for 1 million runs.

\begin{table}[ht]
%\vspace{1.5cm}
    \centering
    \begin{tabular}{|c|c|c|c|}
        \hline
        Dataset & \makecell{Training \\ Time (s)} & Prediction Time (s) & \makecell{Accepted \\ Steps}  \\\hline
        150 & 0.1082 & 294.519 & 1.21$\%$ \\
        250 & 0.0634 & 288.165 & 2.02$\%$ \\
        400 & 0.0612 & 286.410 & 2.07$\%$ \\
        600 & 0.1832 & 294.135 & 2.70$\%$ \\
        800 & 0.1972 & 404.744 & 2.45$\%$ \\
        1000 & 0.1648 & 403.346 & 2.09$\%$ \\
        1200 & 0.2896 & 388.484 & 3.94$\%$ \\
        HADES & 1.843 & 362.654 & 15.61$\%$ \\
        \hline
    \end{tabular}
    \caption{Time trials for DNN emulators for all FOPI and HADES datasets.}
    \label{DNNTT}
\end{table}

\begin{table}[ht]
    \centering
    \begin{tabular}{|c|c|c|c|}
        \hline
        Dataset & \makecell{Training \\ Time (s)} & Prediction Time (s) & \makecell{Accepted \\ Steps}  \\\hline
        150 & 60.785 & 6.004 & 1.10$\%$ \\
        250 & 58.892 & 6.197 & 2.21$\%$ \\
        400 & 59.491 & 6.226 & 2.41$\%$ \\
        600 & 327.798 & 6.134 & 2.97$\%$ \\
        800 & 64.449 & 6.278 & 2.58$\%$ \\
        1000 & 66.501 & 6.302 & 2.18$\%$ \\
        1200 & 1140.1 & 12.739 & 4.27$\%$ \\
        HADES & 1882.5 & 12.659 & 15.62$\%$ \\
        \hline
    \end{tabular}
    \caption{Time trials for symbolic-regression emulators for all FOPI and HADES datasets.}
    \label{SymRegTT}
\end{table}
\newpage
\bibliographystyle{nst}
\bibliography{refs}

\end{document}